\def \A {{\bm A}}

\def \At {A_{\theta}}
\def \Ap {A_{\phi}}
\def \fxt {{\bm f}({\bm x},t)}
\def \f {{\bm f}}

\def \AAA {{\bm A}}
\def \WW {{\bm W}}
\def \u {{\bm U}}

\def \B {{\bm B}}

\def \j {{\bm J}}
\def \k {{\bm k}}

\def \grad {{\bm \nabla}}
\def \curl {{\bm \nabla} \times}
\def \dive {{\bm \nabla}\cdot}
\def \lap {\nabla^2}
\def \delt {\partial_t}
\def \Dt {D_t}

\newcommand{\bra}[1]{\langle #1\rangle}
\newcommand{\meanBB}{\overline{\mbox{\boldmath $B$}}{}}{}
\newcommand{\dd}{{\rm d} {}}

\def \Rey  {\mbox{Re}}
\def \Rm  {\mbox{Re}_{\rm M}}

\def \cs  {c_{\rm s}}
\def \kf  {k_{\rm f}}

\def \tl {\tau}
\def \teta {\tau_{\eta}}
\def \lf {\ell_{\rm f}}

\def \Urms {U_{\rm rms}}
\def \Ek {E_{\rm K}}
\def \Em {E_{\rm M}}

\def \Hk {H_{\rm K}}
\def \Rey  {\mbox{Re}}
\def \Rm  {\mbox{Re}_{\rm M}}

\def \cs  {c_{\rm s}}
\def \kf  {k_{\rm f}}

\def \tl {\tau}
\def \teta {\tau_{\eta}}
\def \lf {\ell_{\rm f}}

\def \Urms {U_{\rm rms}}
\def \Ek {E_{\rm K}}
\def \Em {E_{\rm M}}

\def \Hk {H_{\rm K}}
\newif\iffigs
\figstrue
\iffigs
\fi
\def\drawing #1 #2 #3 {
\begin{center}
\setlength{\unitlength}{1mm}
\begin{picture}(#1,#2)(0,0)
\put(0,0){\framebox(#1,#2){#3}}
\end{picture}
\end{center} }
\documentclass[preprint2]{emulateapj}
\usepackage{lscape,rotating}
\usepackage{amsmath,amssymb}
\usepackage{times}
\usepackage{graphicx}
\usepackage{fancyhdr}
\usepackage{bm}
\graphicspath{{./fig/}{./png/}{./plots/}}

\shorttitle{Dynamos in spherical geometry}
\shortauthors{D. Mitra et al.}

\begin{document}

\title{Turbulent dynamos in spherical shell segments of varying geometrical extent}
\author{Dhrubaditya Mitra \& Reza Tavakol}
\affil{Astronomy Unit, School of Mathematical Sciences, Queen Mary 
      University of London, Mile End Road, London E1 4NS, UK}
\email{dhruba.mitra@gmail.com}
\author{Axel Brandenburg}
\affil{NORDITA, Roslagstullsbacken 23, SE-10691 Stockholm, Sweden}
\email{brandenb@nordita.org}
\author{David Moss}
\affil{School of Mathematics, University of Manchester, Oxford Road, 
       Manchester M13 9PL, U.K.}
\email{David.Moss@manchester.ac.uk}
\begin{abstract}

We use three-dimensional direct numerical simulations of the helically 
forced magnetohydrodynamic equations in spherical shell segments
in order to study the effects of changes in the geometrical shape 
and size of the domain on the growth and saturation of large-scale 
magnetic fields. We inject kinetic energy along with kinetic helicity in 
spherical domains via helical forcing using Chandrasekhar-Kendall functions.
We take perfect conductor boundary conditions for the magnetic field
to ensure that no magnetic helicity escapes the domain boundaries. 
We find dynamo action giving rise to magnetic fields at scales larger
than the characteristic scale of the forcing. The magnetic energy exceeds the 
kinetic energy over dissipative time scales, similar to that seen earlier
in Cartesian simulations in periodic boxes. As we increase the size of 
the domain in the azimuthal direction we find that the nonlinearly 
saturated magnetic field organizes itself in long-lived cellular 
structures with aspect ratios close to unity.
These structures tile the domain along the azimuthal direction, thus 
resulting in very small longitudinally averaged magnetic fields for large 
domain sizes. The scales of these structures are determined by the smallest 
scales of the domain, which in our simulations is usually the radial scale.
We also find that increasing the meridional extent of the domains produces 
little qualitative change, except a marginal increase in the large-scale field.
We obtain qualitatively similar results in Cartesian domains with 
similar aspect ratios.
\end{abstract}


\section{Introduction}

A fundamental question in solar and stellar physics concerns the
generation of large-scale magnetic fields in convective spherical 
shells through dynamo action, which occurs on dynamical time scales.
A great deal of effort has gone into
understanding this question by using direct three dimensional
magnetohydrodynamic (MHD) simulations, in Cartesian domains
with forced and convective turbulence as well as in spherical domains.
These studies can be divided into four broad 
groups. The first consists of helically forced turbulence simulations
in Cartesian domains, see e.g., \cite{B01,BD01}.
These simulations in general show large-scale magnetic fields when
periodic or perfect conductor boundary conditions are used, but only 
growing
on dissipative time scales, which makes them not directly 
relevant to solar and stellar situations.
With more realistic open boundary conditions and in presence of shear,
large-scale magnetic fields are known to develop on dynamical
time scales \citep{B05}. 
The second group comprises simulations of turbulent
convection in Cartesian coordinates, which have recently
shown large-scale magnetic fields \citep{KKB08,HP08}.
Thirdly, forced incompressible turbulence simulations 
in full spheres, mostly relevant to 
planetary dynamos,  have been carried out by \cite{MM06,MMT07}.
Finally, there is an increasing body of work employing simulations of 
MHD turbulence in spherical rotating shells with convection using
the incompressibility constraint with either 
Boussinesq approximations \citep{GM81,G83} or anelastic approximations
\citep{GG81,GG82,G84,G85,MET00,BT02,BMT04,BMT06,BBBMNT08}.
These simulations produce mainly small-scale magnetic fields and only
insignificant large-scale magnetic fields
with parameters relevant to the solar and stellar settings. 
Relatively stronger large-scale (global)
magnetic fields have, however, been found in 
rapidly rotating shells \citep{BBBMNT08}.
Also, it has recently been shown
that in simulations of fully convective stars
the energy in the longitudinally averaged magnetic field 
can become locally comparable to the kinetic energy \citep{BR08}.

In the present paper we attempt to bridge the gap between studies
in Cartesian and spherical shell domains by
solving the MHD equations in wedge-shaped domains of spherical shells
with helical forcing. In particular we study the effects of
shape and size of the computational 
domain on the growth and saturation of the large-scale magnetic field.
Spherical wedge geometries in principle provide an advantage
in terms of computational resources over both the Cartesian boxes and 
spherical shell geometries usually employed in MHD simulations 
in that they strike a reasonable compromise 
between the requirements for spatial resolution and globality.
In other words, our choice of spherical wedge domains allows in principle
higher  absolute spatial resolution (i.e., higher number of 
grid points per unit length), thus potentially allowing larger 
magnetic Reynolds numbers
(defined later) to be attained, whilst retaining some globality.
Alternatively, at a given resolution, we can achieve simulations in
a number of wedge domains, or in one domain for much longer
time, for the cost of one simulation in a full spherical shell --
this is the approach we adopt here.

In this paper we make a number of assumptions that are motivated by the
desire to understand the basic concepts of dynamo saturation in spherical
geometries instead of providing a realistic model of the solar dynamo.
Specifically, we consider here the case of homogeneous turbulence with
perfectly conducting boundary conditions so as to make contact with
corresponding earlier work in Cartesian domains.
The physically more relevant case of open boundary conditions with an 
equator and differential rotation or shear will be postponed to future 
work.

The paper is organized as follows.
In Sect.~\ref{model} we briefly describe
the details of our model and the code used.
Sect.~\ref{results} contains our results, where
for the sake of clarity, we present the results
concerning the effects of increasing the domain in the azimuthal and 
meridional directions separately.
Sect.~\ref{conclusion} contains our conclusions. 
Finally, Appendices A and B contain the details of the
helical forcing used, and our extension of the
\textsc{Pencil Code}\footnote{
\url{http://www.nordita.org/software/pencil-code}.}
to non-Cartesian coordinate systems,
respectively.

\section{The model}
\label{model}
We solve numerically the magnetohydrodynamic equations 
for the velocity $\u$, the logarithmic density $\ln\rho$,
and the vector potential $\A$, given by
\begin{equation}
\Dt\u = -c_{\rm s}^2\grad\ln\rho + \frac{1}{\rho}\j\times\B 
 + \bm{F}_{\rm visc} + \f,
\label{mhd1}
\end{equation}
\begin{equation}
\Dt\ln\rho = -\grad\cdot\u,
\end{equation}
\begin{equation}
\delt\A =  \u\times\B + \eta\lap\A ,
\end{equation}
where $\bm{F}_{\rm visc}=(\mu/\rho)(\lap\u + \frac{1}{3}\grad\dive\u)$
is the viscous force, $\mu$ is the dynamic viscosity, 
$\B = \curl \A$ is the magnetic field,
$\j = \curl \B/\mu_0$ is the current density,
$\mu_0$ is the vacuum permeability,
$c_{\rm s}^2$ is the velocity of sound in the medium, $\rho$ is the density,
$\eta$ is the magnetic diffusivity,
and $D_t \equiv \delt + \u\cdot\grad$ is the advective derivative.
Here $\fxt$ is an external random helical forcing (the details 
of which are given in Appendix~\ref{helf}), satisfying the condition,
\begin{equation}
\f\cdot\curl\f \ge 0\/,
\label{hel}
\end{equation}
in order to ensure positive helicity injection over the entire sphere.
Such a model is reminiscent of constant $\alpha$ effect spheres that were
studied in the early days of mean-field dynamo theory \citep{KS67}.
Similar cases relevant to planetary dynamos have also been studied
recently by direct numerical simulations \citep{MM06,MMT07}.

\begin{figure}[th]
\iffigs
\includegraphics[width=\linewidth]{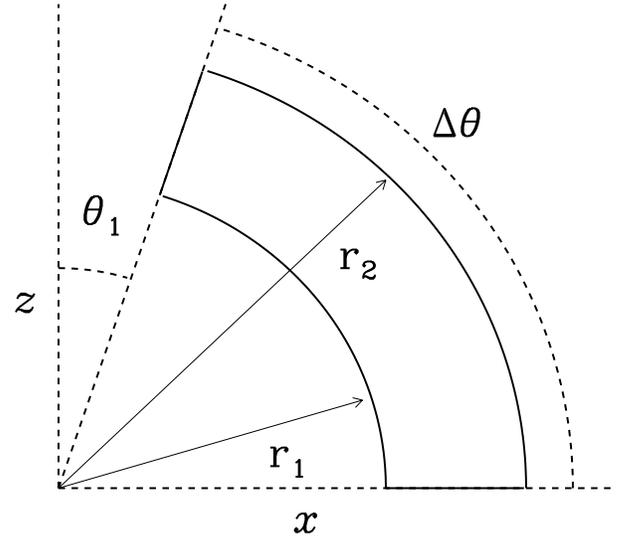}
\else\drawing
  50 30 {  } \fi
\caption{\small Schematic representation of the 
meridional plane of our spherical wedge computational domain.
We also define $\theta_2$ to be the angle that the other azimuthal 
boundary makes with the polar axis; in this Figure and throughout 
this paper $\theta_2=\pi/2$.}
\label{wedge}
\end{figure}

A sketch of the meridional cross-section of a typical wedge-shaped domain
used in our simulations is given in Fig.~\ref{wedge}.
We confine ourselves to simulations in the northern hemisphere.
However, because there is no rotation, the choice of the coordinate axis
(and hence of the equator) is arbitrary.
Therefore the physical conditions are the same on
either side of the equator.
The code used for our computations is the \textsc{Pencil Code}
developed by \cite{BD02b} in Cartesian coordinates.
We have extended the code to allow simulations in spherical coordinates.
This was facilitated by the fact that the \textsc{Pencil Code} was already written in
a non-conservative form, which allowed the curvilinear
coordinates to be implemented by replacing all partial derivatives 
by covariant derivatives, see Appendix ~\ref{noncart} for further details.

Guided by the convection zone of the Sun, in the majority
of our computations the radial extent of our domain is chosen 
to be $0.7 \le r\le 1.0$. We use perfect conductor boundary 
conditions for the magnetic field to ensure that no magnetic 
helicity escapes the domain boundaries. 
In Cartesian domains \citep{B01} this is often achieved
by assuming periodic boundary conditions across the boundaries.
In our spherical case this translates to the normal component of
the magnetic field $\B$ being
continuous (and hence zero) across the boundary. This implies that
the tangential components of the magnetic vector potential  $\A$
must be zero at the boundary. We are free to choose the 
boundary condition for the normal component. Guided by this,
we make the following choices at the four boundaries of our domain
\begin{equation}
\At=\Ap=\frac{d A_r}{dr}=0\quad\mbox{(on $r=r_1$)},
\label{bc_r1}
\end{equation}
\begin{equation}
\At=\Ap=A_r=0\quad\mbox{(on $r=r_2$ and $\theta=\theta_2=\pi/2$)},
\label{bc_r2}
\end{equation}
\begin{equation}
A_r=\Ap=\frac{d \At}{d\theta}=0\quad\mbox{(on $\theta=\theta_1$)}.
\label{bc_t1}
\end{equation}
There is no particular reason for using $A_r=0$ on $r=r_2$ and not on
$r=r_1$, and we emphasize that the condition on the normal component of
$\bm{A}$ is of no significance for $\bm{B}$ itself.
We use stress-free boundary conditions for the velocity at all these four
boundaries and periodic boundary conditions for all the variables along 
the azimuthal direction.

\section{Results}
\label{results}
Our principle aim in this paper is to study the growth and saturation of
large-scale magnetic field in spherical wedge domains. In particular we
study the effects of changes in the shape and the size of the domain on the 
resulting large-scale fields. For the sake of clarity we
do this by studying the effects of 
increasing the domain extent in the $\theta$ and
$\phi$ directions in turn. We also briefly 
look at the role of the radial extent of the computational domain.
Given that the size of our simulation domain is different along different directions
in different runs, we in general have three different length scales, 
$L_{\rm r} \equiv r_2 - r_1$, $L_{\theta} \equiv r_2(\theta_2-\theta_1)$ and 
$L_{\phi} \equiv r_2\sin\theta_2(\phi_2-\phi_1)$,
corresponding to the sizes of the domain in 
the $r$, $\theta$ and $\phi$ directions respectively.
As an estimate of the characteristic Fourier mode of forcing we use
$\kf=W_{\rm rms}/U_{\rm rms}$, where $W_{\rm rms}=\langle\WW^2\rangle^{1/2}$
is the rms value of the vorticity, $\WW \equiv \nabla \times \u$,
and $U_{\rm rms}$ is the rms velocity.
Here, angular brackets denote volume averages.
The characteristic length scale of forcing is defined to be
$\lf \equiv 2\pi/\kf$. 
We then define the fluid Reynolds number, magnetic Reynolds number 
 and the turnover time
as $\Rey=\Urms/\nu\kf$, $\Rm=\Urms/\eta\kf$ and 
$\tl\equiv(\Urms\kf)^{-1}$ respectively. Here, $\nu$ is the kinematic
viscosity given by $\nu = \mu/\rho_0$ where $\rho_0$ is the
initial density (which is equal to the mean density throughout,
noting that the mass in the volume is conserved).
In all our runs $\tl$ is nearly the same and varies from  
$0.6\cs/r_2$ (run~{\tt S7}) to $0.9\cs/r_2$ (run~{\tt S1}).
As the dynamo we study is resistively limited \citep[cf.][]{B01},
the time scale of saturation is the dissipative time scale
$\teta=\eta\k_1^2$ which is used to non-dimensionalize the time.
$\teta=\eta\k_1^2$ which we use to normalize the time axes of
our plots. Here $k_1$ is the wavenumber corresponding to the 
smallest length scale in our domain, i.e., $k_1\equiv2\pi/L_r$
for most of our runs.
The helical nature of the velocity field is characterized by
$H_{\rm K}=\langle \WW \cdot \u \rangle/(W_{\rm rms}\Urms)$.
We start our simulations with
a zero velocity field and a Gaussian random magnetic vector potential
such that the amplitude of the magnetic field is of the order of
$10^{-6}$ in units of $(\rho_0\mu_0)^{1/2}\cs$.

The growth and saturation of the magnetic dynamo is monitored by
the total magnetic energy, $\Em = \bra{\B^2}/2\mu_0$,
and the kinetic energy, $E_{\rm K} = \bra{\rho\u^2}/2$.
We define the large-scale (or mean) magnetic field using longitudinal averaging,
\begin{equation}
{\overline {\bf B}}(r,\theta,t) \equiv \frac{1}{2\pi} \int \bm{B}\,{\rm d}\phi,
\end{equation}
over the extent of the domain.
The total energy in the large-scale magnetic field is then defined by
$E_{\rm LS}(t) = \bra{\meanBB^2}/2\mu_0$.

A measure of the level of turbulence in our simulations is the Reynolds number, 
$\Rey$, given in Table~\ref{runpars},
which is about $5$ in all the runs except run {\tt S7} in which case it is $2$. 
For all practical purposes there is essentially no inertial range in the 
spectrum of the fluid obtained from our runs.
\begin{deluxetable}{ccccccccc}
\tablecaption{Parameters of the spherical runs.\label{runpars}}
\tablehead{
\colhead{Runs} & \colhead{Grid} & \colhead{$L_{\theta}$} & 
         \colhead{$L_{\phi}$} & \colhead{$\lf/L_r$} &
          \colhead{$\Rey$} & \colhead{$\Rm$} & \colhead{$H_{\rm K}$} &
           \colhead{$\lambda$} }
\startdata
{\tt S1}& $32\times32\times32$ & $0.1\pi$ & $0.1\pi$ &
              $0.5$ &  $5$ & $14$ & $0.66$ & $0.08$\\
{\tt S2} & $32\times32\times128$ & $0.1\pi$ & $\pi/2$ & 
          $0.4$ &  $5$ & $12$ & $0.74$ & $0.10$  \\
{\tt S3} & $32\times32\times256$ & $0.1\pi$ & $\pi$ &
                 $0.4$ &  $5$ & $12$ & $0.74$ & $0.10$\\
{\tt S4} & $32\times64\times32$ & $0.2\pi$ & $0.10\pi$ & 
               $0.5$ &  $5$ & $12$ & $0.65$ & $0.10$\\
{\tt S5} & $32\times256\times32$ & $85^\circ$ & $0.1\pi$ & 
               $0.4$ &  $5$ & $12$ & $0.73$ & $0.14 $\\
{\tt S6} & $32\times64\times128$ & $0.2\pi$ & $\pi/2$ & 
                $0.4$ &  $5$ & $11$ & $0.79$ & $0.10$\\
{\tt S7} & $32\times32\times64$ & $0.1\pi$ & $\pi/4$ &
                $0.2$ &  $2$ & $4$ & $0.79$ & $0.10$\\
\enddata
\end{deluxetable}
The summary of the runs together with their domain sizes, resolutions and
other relevant parameters are
given in Table \ref{runpars}. 
In the following subsections we summarize the results of our simulations as
the domain sizes in the azimuthal and meridional directions, $L_{\phi}$ and
$L_{\theta}$ respectively, are changed separately.
\subsection{Initial growth phase}
\begin{figure}[ht]
\iffigs
\includegraphics[width=\linewidth]{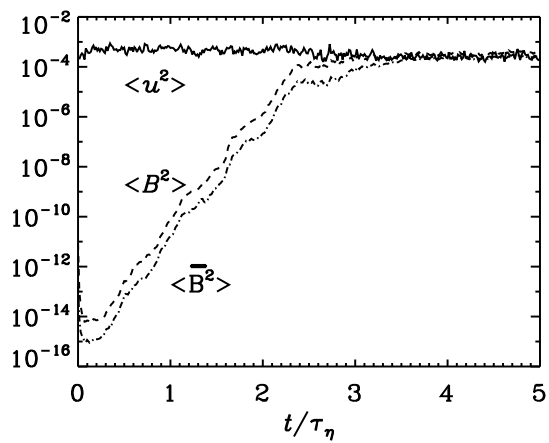}  
\else\drawing
  50 30 {  } \fi
\caption{\small Evolution of $\langle\u^2 \rangle$ (continuous),
$\langle\B^2 \rangle$ (dashed) and
$\langle \overline{\B}^2\rangle$ (dash-dotted)
during early times from run {\tt S1}. Similar
exponential growth of the magnetic energy is seen in all the other runs.}
\label{energy_shorttime_run1}
\end{figure}

We first summarize our results concerning the
growth phase of the dynamos. We begin with the run with the smallest
domain size, i.e., {\tt S1}; see Fig.~\ref{energy_shorttime_run1}.
As can be seen the magnetic energy starts growing exponentially 
from $t \approx 0.2 \teta$ and the total magnetic energy
reaches the level of the kinetic energy at $t\approx 3\teta$.
This is true of all our runs, since they all start with the same
initial field strength, and they all have the same growth rate which,
in turn, is proportional to $\Urms\kf$, which is also the same for all runs.
The growth rate during this exponential growth phase is given by 
\begin{equation}
\lambda(t)=\frac{\dd}{\dd t} \ln \langle\B^2 \rangle_{\rm lin}^{1/2},
\end{equation}
which is about $0.1$ for all the runs performed here; see Table~\ref{runpars}.
In all cases $\Ek$ decreases (i.e.\ it is quenched) 
after $\Em$ reaches saturation. 
We note that even after reaching saturation the field keeps growing 
somewhat, similar to what has been seen earlier in Cartesian domains 
with periodic boundary conditions \citep{B01}, or with perfectly conducting
boundaries \citep{BD02b}.
Both $\Ek$ and $\Hk$ decrease slightly (by less than 10\%) after saturation is 
reached for runs {\tt S2}, {\tt S3}, and {\tt S6}. 
For other runs the $\Hk$ decreases a little
more (by factors from about $0.8$ to $0.6$).
\begin{figure}[ht]
\iffigs
\includegraphics[width=\linewidth]{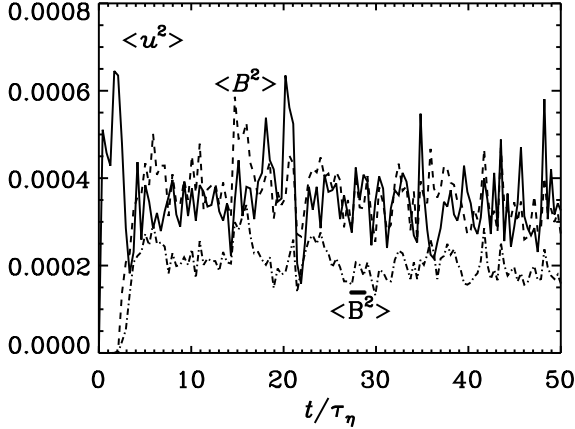}  
\else\drawing
  50 30 {  } \fi
\caption{\small Evolution of kinetic energy $\langle\u^2 \rangle$ (continuous), 
magnetic energy $\langle\B^2 \rangle$ (dashed) and energy in the 
large-scale magnetic field $\langle \overline{\B}^2\rangle$ (dash-dotted) 
during late times from run {\tt S1}. The saturated value of the energy in 
 the large-scale magnetic field is comparable to the kinetic energy.}
\label{energy_s1}
\end{figure}
\subsection{Formation of large-scale magnetic field}
\begin{figure}[h]
\iffigs
\includegraphics[width=\linewidth]{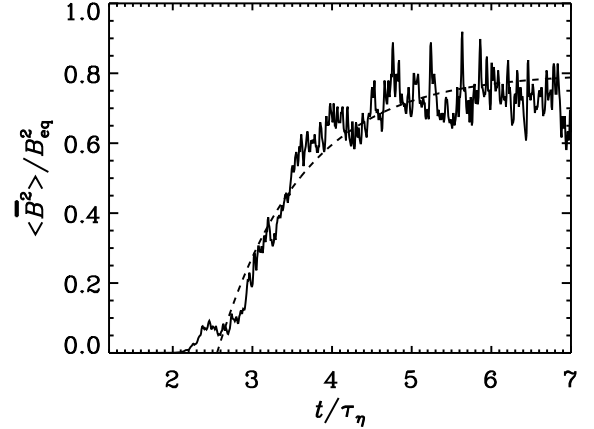}  
\else\drawing
  50 30 {  } \fi
\caption{Late saturation behavior of the mean field $\overline{\B}$
for the run {\tt S1} compared with the prediction given by Eq.~(\ref{efit})
(dashed).
}
\label{bsat}
\end{figure}

For the smallest domain chosen here, i.e.\ {\tt S1} which is closest 
to a cube ($L_r\approx L_\theta\approx L_\phi$),
we obtain results that are very similar to those
found earlier from Cartesian simulations \citep{B01}.
The large-scale magnetic field grows, reaches a value close to
equipartition and then shows a slow saturation on dissipative 
time scales, see Fig.~\ref{energy_s1}. As we are using
perfectly conducting boundary conditions the growth of 
the large-scale magnetic field 
is limited by the decay of small-scale magnetic helicity.
This has been used to model the saturation of 
the magnetic energy of the large-scale field \citep{B01},
\begin{equation}
\frac{{\overline B}^2}{B_{\rm eq}^2}
=\frac{\epsilon_{\rm f}k_{\rm f}}{\epsilon_{\rm m}k_{\rm m}}
\left[1-e^{-2\eta k_m^2(t-t_{\rm sat})}\right].
\label{efit}
\end{equation}
Here $t_{\rm sat}$ is the approximate time when the small-scale field
has saturated, and $k_m$ is a new effective wavenumber
which is related to $k_1$ and is treated here as a fit parameter
that is chosen to match the simulation result.
We obtain $k_m\approx0.7k_1$.
Here $B_{\rm eq}^2$
corresponds to the kinetic energy density,
so $B_{\rm eq}^2/\mu_0=\bra{\rho u^2}$,
and is approximately equal to the energy in the small-scale magnetic field.
Expression (\ref{efit}) fits the data from our simulations quite well
as shown in Fig.~\ref{bsat}. 
  
\begin{figure}[h]
\iffigs
\includegraphics[width=\columnwidth]{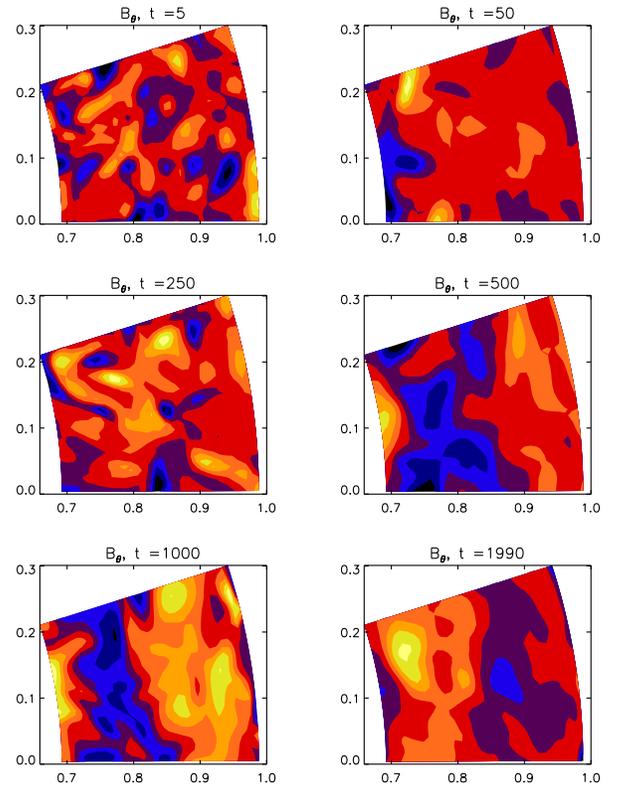}  
\else\drawing
  50 30 {  } \fi
\caption{Contour plots of $B_{\theta}$ in the equatorial
plane of the domain in {\tt S1} at different times showing the
gradual establishment of a large-scale magnetic field.
Time is here given in units of $R/\cs$.
}
\label{bphi_equi_run1}
\end{figure}

\begin{figure}[h]
\iffigs
\includegraphics[width=\columnwidth]{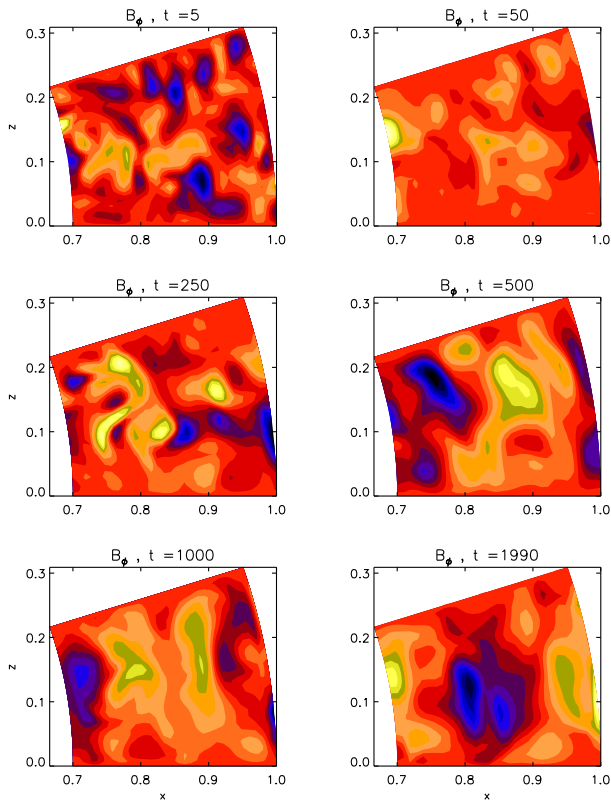}  
\else\drawing
  50 30 {  } \fi
\caption{Contour plots of $B_{\phi}$ in the meridional
plane of the domain in {\tt S1} at different time.
Time is here given in units of $R/\cs$.
}
\label{br_mean_t_run1}
\end{figure}

The evolution of the large-scale magnetic
field follows closely the evolution of the total magnetic
field after the time when the amplitude
of the large-scale field has become steady.
The growth of large-scale structures for this run (run {\tt S1})
in the equatorial plane and the meridional plane are shown in 
Figs.~\ref{bphi_equi_run1} and ~\ref{br_mean_t_run1} respectively. 
Large-scale structures in the contour plots of magnetic field in the 
equatorial plane appear as early as $ t \approx 500$ (about $6 \teta$)
and at late times they encompass the whole azimuthal extent of the domain. 

\subsection{Effects of increasing the azimuthal extent}
\label{azimuth}
\begin{figure}[ht]
\iffigs
\includegraphics[width=\linewidth]{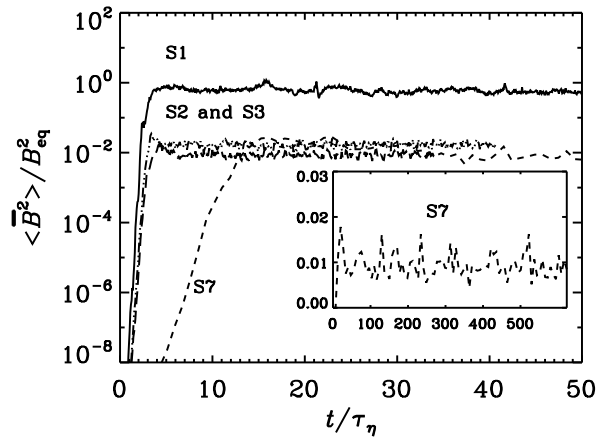}  
\else\drawing
  50 30 {  } \fi
\caption{Normalized energy in the large-scale magnetic field versus time 
for the runs {\tt S1}, {\tt S2}, {\tt S3} and {\tt S7}. 
As can be seen, as the domain size increases in the $\phi$-direction  
the field decreases (see also Fig.~\ref{el-phi}). The
inset shows the same plot but in linear scale for the run {\tt S7} which 
was run more than 10 times longer than the other cases.}
\label{phi-depen}
\end{figure}

To proceed, we begin by increasing the domain size 
in the $\phi$ direction, while keeping the $\theta$ and $r$ 
dimensions fixed, and increase thereby the aspect ratio.
The initial growth phase remains practically unchanged as we 
change the extent of our domain.
We find that the large-scale magnetic field 
decreases as we go to larger domains (by increasing $L_{\phi}$),
as can be seen in Fig.~\ref{phi-depen}.

\begin{figure}[h]
\iffigs
\includegraphics[width=\linewidth]{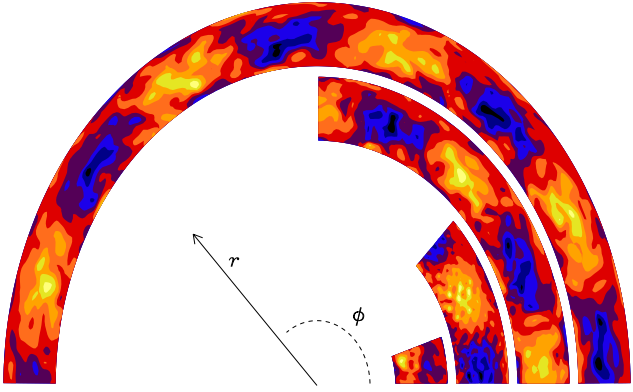}  \\
\else\drawing
  50 30 {  } \fi
\caption{Contour plots showing the typical structure of 
the magnetic field in the equatorial plane as the $\phi$ extent
of the domain is increased. 
From top to bottom, plots of the runs {\tt S3},{\tt S2}, {\tt S7} and {\tt S1}.}
\label{phi-largescale}
\end{figure}

\begin{figure}[h]
\iffigs
\includegraphics[width=\columnwidth]{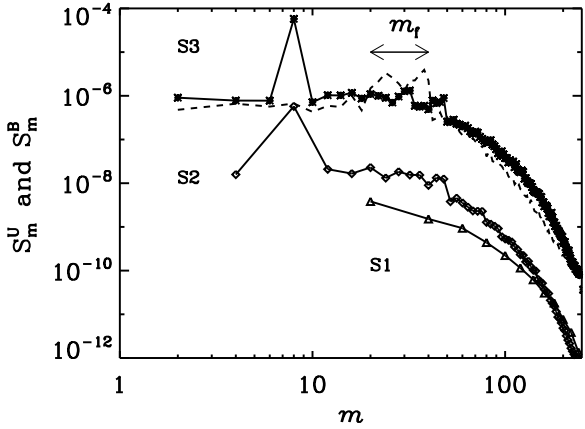}  
\else\drawing
  50 30 {  } \fi
\caption{\small Kinetic and magnetic energy spectra $S^{\rm U}_m$ (dashed line)
and $S^{\rm B}_m$ (continuous line) from runs {\tt S1}, {\tt S2},
and {\tt S3}. The range of scales being forced are shown within the two
arrowheads. For clarity the spectrum for run {\tt S1} and {\tt S2} are
 multiplied by a factor of $10^{-4}$ and $10^{-2}$ respectively. 
}
\label{spec-phi}
\end{figure}

To understand the reason for this decrease in the large-scale 
magnetic field we present contour plots of the $\theta$ component
of magnetic field in the equatorial plane for four different 
domain sizes at later times (see Fig.~\ref{phi-largescale}). Notice
that as we increase our domain size `cell-like' structures are developed 
along the azimuthal direction, with aspect ratios
close to unity. Their typical length scale corresponds to and seems to be 
determined by
the smallest dimension of our domain, which here is the radial extent. 
We checked this by performing a run with half the radial
extent and found that the characteristic horizontal scale
of the cell structures is decreased accordingly, to half the original value.
We also find that the length scale of these cell structures
does not depend on the 
forcing length scale. We verified this by changing the forcing
length scale along the radial direction and found that this
does not change these cell structures. 
The length scale of the cells is also larger than the characteristic
length scale of the velocity. This is best described 
using Fourier transform along the azimuthal (periodic) direction,
\begin{equation}
\hat\u_m(r,\theta) = \int \u(r,\theta,\phi)\exp(im\phi)\,{\dd\phi\over2\pi},
\end{equation} 
\begin{equation}
\hat\B_m(r,\theta) = \int \B(r,\theta,\phi)\exp(im\phi)\,{\dd\phi\over2\pi}.
\end{equation} 
We can define the averaged spectra of these Fourier transformed quantities as,
\begin{equation}
S^{\rm U}_m=\bra{|\hat \u_m(r,\theta)|^2}_{r\theta},\quad
S^{\rm B}_m=\bra{|\hat \B_m(r,\theta)|^2}_{r\theta},
\end{equation} 
where the subscript $r\theta$ denotes meridional averaging.
Note that $\sum S^{\rm U}_m=\langle\u^2\rangle$ and
$\sum S^{\rm B}_m=\langle\B^2\rangle$.
We plot in  Fig.~\ref{spec-phi} both $S^{\rm B}_m$ and $S^{\rm U}_m$ for
the runs {\tt S1}, {\tt S2} and {\tt S3}, which have
azimuthal extents $\pi/10$, $\pi/2$ and $\pi$ respectively.
We find that the peak in the spectrum of the magnetic field occurs 
at the same $m$ for runs {\tt S2} and {\tt S3},
showing that the typical characteristic 
scale of the periodic structures does not
depend on the  $\phi$ extent of our domain. 
Note also that the typical forcing scale is clearly smaller than this
(corresponding to $m\approx20$--$40$).

An important question regarding these structures, and 
hence the resulting large-scale magnetic fields, is 
whether these periodic structures are transient
and may later merge to form structures encompassing
the whole domain similar to the run {\tt S1} with the smallest domain size.
The characteristic time
scale over which structures encompassing the whole domain form in {\tt S1}
is about $6 \teta$.
In Cartesian simulations with magnetically closed boundaries the saturation
time is inversely proportional to the square of the relevant domain size.
Thus, by analogy, if the $\phi$ extent is doubled  the time scale $\sim 6\teta$
would become $\sim 24 \teta$.
Similarly it would take even longer for such structures to form in the runs with
bigger domain size. We have studied a run -- run {\tt S7} --  
in which the azimuthal extent of the domain
is twice that of {\tt S1} and have run this simulations up 
to $500 \teta$, without 
finding any evidence of cells merging. 
This suggests that the periodic structures that we
observe are at least as long-lived as the duration of our longest runs. 

\begin{figure}[ht]
\iffigs
\includegraphics[width=\linewidth]{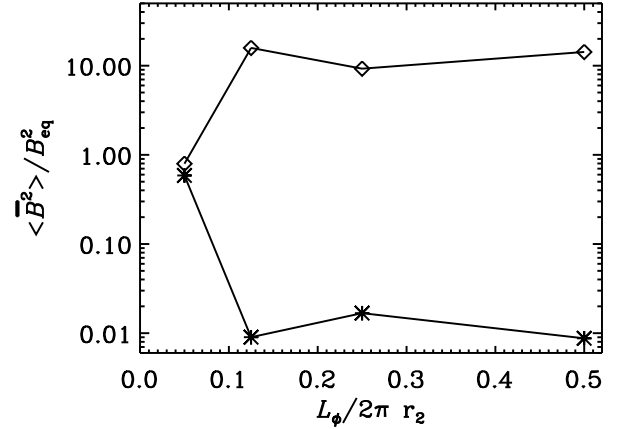}  
\else\drawing
  50 30 {  } \fi
\caption{\small Two different measures to estimate the energy in the 
large-scale magnetic field plotted against $L_{\phi}$, for four different
runs having different domain sizes along the azimuthal direction. 
In one case the large-scale magnetic energy is
estimated by longitudinal average (denoted by $\ast$ in the plot), in the
other case (denoted by $\diamond$ in the plot) it is estimated by the 
magnitude of the peak of $S_m^B$.}
\label{el-phi}
\end{figure}

To summarize, 
our simulations show that the characteristic scale of the large-scale 
magnetic fields found in our simulations
is about the scale of the radial extent of our domain. 
Hence, as we increase $L_{\phi}$ an increasing number of 
periodic structures appear along the azimuthal direction,
which, in the largest domain we have used, is about $10$ times
larger than the radial direction.
Therefore the large-scale magnetic field, defined as a longitudinal average,
gives a very small contribution in the runs with larger domains. 
Note that, with this definition, the energy of the large-scale
magnetic field corresponds to the energy in the (axisymmetric) $S^B_0$ mode.
From the plot of the spectrum 
we note that most of the magnetic energy is actually
concentrated at $m=8$, which is the scale of a cell,
and this mode indeed shows super-equipartition.  
Also note that this mode corresponds to length scales larger than the
scale of forcing, which corresponds to $m \approx 20 - 40$. 
Hence, instead of using the longitudinal average to calculate the energy in the
large-scale magnetic field we can use the energy in the mode $m=8$ of 
the meridionally averaged spectrum of the magnetic field.
A comparison between these two methods of calculation of energy
in the large-scale magnetic field is shown in Fig.~\ref{el-phi}. 
Note that as $L_{\phi}$ is increased the large-scale 
magnetic energy measured by the averaged spectrum of the magnetic field 
remains practically constant. 

\subsection{Effects of increasing the meridional extent}
\label{merid}
\begin{figure}[h]
\iffigs
\includegraphics[width=\columnwidth]{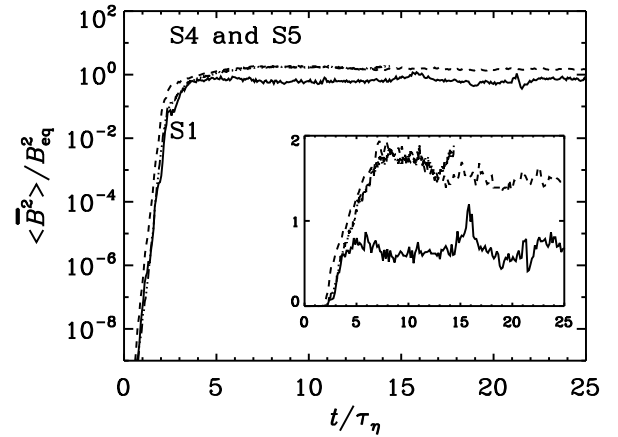} \\ 
\else\drawing
  50 30 {  } \fi
\caption{Normalized energy in the large-scale magnetic field versus 
time for three different runs {\tt S1}, {\tt S4}, and {\tt S5}. The
inset shows the same plot but in linear scale.}
\label{theta-depen}
\end{figure}

\begin{figure}[h]
\iffigs
\includegraphics[width=\columnwidth]{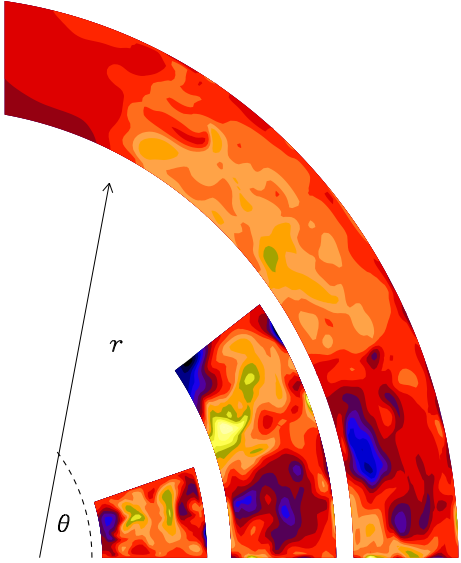} 
\else\drawing
  50 30 {  } \fi
\caption{Meridional cross-sections of $B_\phi$ after saturation has
been reached. From left to right: {\tt S1}, {\tt S4},
and {\tt S5}
}
\label{theta-largescale}
\end{figure}

Next we study the effects of increasing the domain size by increasing the
 meridional extent. We find that, as we increase the domain size along the 
$\theta$ direction, the large-scale magnetic field shows marginal increase;
see Fig.~\ref{theta-depen}. Contour plots of the toroidal 
component of the magnetic field in the meridional plane for the runs {\tt S1},
{\tt S4} and {\tt S5} are shown in Fig.~\ref{theta-largescale}.
Note that, as the domain is increased in the $\theta$ direction,
the field structure at low latitudes is largely unchanged, while
new weaker fields are added at high latitudes.
However, the high latitudes contribute relatively little to the
volume average, so the magnetic energy is only marginally
increased. 
As noted above, our use of the term high latitudes is defined by our 
arbitrary choice of the coordinate axis -- see Fig.~\ref{wedge}. 
However, once such a choice is  made, the field can only develop subject to 
the constraints imposed by the geometry of the computational domain.

\begin{figure}[h]
\iffigs
\includegraphics[width=\columnwidth]{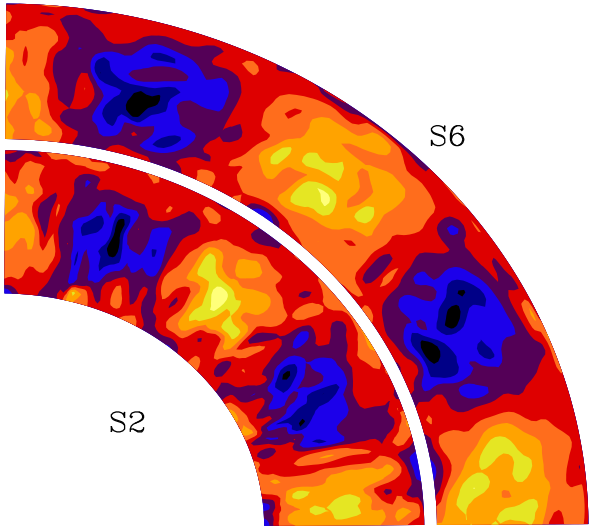} 
\else\drawing
  50 30 {  } \fi
\caption{Equatorial cross-sections of $B_\theta$ after saturation has
been reached from runs {\tt S2} and {\tt S6},
}
\label{azi-meri-largescale}
\end{figure}

Again, we have checked that the characteristic scale of the cells is
determined by the smallest of the three dimensions of our domain,
by performing a simulation in which the meridional extent of the
domain is half that of run {\tt S1}. We find that the resulting
cells again have length scales comparable to the smallest scale of
the domain which in this case is the meridional extent.
Furthermore, we have checked that,
as we increase our domain along the azimuthal extent,
the cell-like structures in our simulations are independent of the
meridional extent of the domain,
provided that the radial scale remains the smallest.
To illustrate this we 
compare contour plots of $B_{\theta}$ in the equatorial plane for the
runs {\tt S2} and {\tt S6} in Fig.~\ref{azi-meri-largescale}. 
These two runs have the same azimuthal extent, 
$L_{\phi}= \pi/2$, but different meridional extents, viz.,
$L_{\theta}=\pi/10$ for
the run {\tt S2} and $L_{\theta}=\pi/4$ for the run {\tt S6}. 
As can be seen the cell-like
structures that appear have the same global features. 
Finally we have checked that the characteristic length scale of cells
is independent of the forcing scale by performing a simulation
in which the characteristic scale of forcing is half that of the
scale of forcing in run {\tt S1}.


\begin{deluxetable}{ccrrrcccr}
\tablecaption{Summary of the runs, including the extents of the
computational domains for the Cartesian runs.
\label{runpars_cart}
}
\tablehead{
\colhead{Runs} & \colhead{Grid} & \colhead{$L_{\rm y}$} & 
            \colhead{$L_{\rm z}$} & \colhead{$\lf/L_{\rm x}$} &
            \colhead{$\Rey$=$\Rm$} & \colhead{$H_{\rm K}$} & 
             \colhead{$\tl$} }
\startdata
{\tt C1} & $32\times32\times32$ &  $6$ & $6$ & $0.5$ &  $16$ & $0.7$ &  $2.6$\\
{\tt C2} & $32\times32\times128$ &  $6$ & $12$ & $0.7$ &  $25$ & $0.6$ & $3.6$ \\
\enddata
\end{deluxetable}
\begin{figure}[ht]
\iffigs
\includegraphics[width=0.95\linewidth]{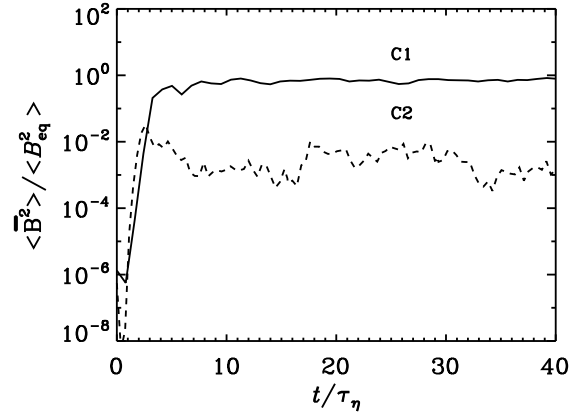}  
\else\drawing
  50 30 {  } \fi
\caption{Normalized energy in the large-scale magnetic field versus time 
for the runs {\tt C1}, {\tt C2}. 
As can be seen as the domain size increases in the $z$-direction, 
the field decreases in a way very similar to the spherical case.
}
\label{tscart}
\end{figure}
\subsection{Cartesian versus spherical: geometry versus aspect ratio}
An important question concerning our results is 
how to differentiate between the effects of geometry (globality) 
and changes in the aspect ratio.  
To answer this question we need to compare our wedge domain simulations with
simulations in Cartesian boxes with appropriate aspect ratios.
To this end we perform two simulations in Cartesian coordinates 
-- runs  {\tt C1} and {\tt C2} -- with aspect ratios one and two
respectively. Relevant parameters for these runs are summarized in 
Table~\ref{runpars_cart}.  
These Cartesian simulations correspond to the runs {\tt S1} and {\tt S7} in
the spherical wedge domains.  The main features of our spherical runs are
also found in the Cartesian runs. In particular, we find that the 
initial (kinematic) growth rate is the same for the runs {\tt C1}
and {\tt C2}. We also observe formation of cell-like structures with
unit aspect ratio in the run {\tt C2}.  
Figure~\ref{tscart} gives a summary of our Cartesian simulations showing plots
of kinetic and  magnetic energy versus time. 
Comparing these results with the corresponding plots for our spherical wedge
runs {\tt S1} and {\tt S7} we observe that the 
decrease in the large-scale magnetic field is 
similar to those in the Cartesian domains, if the aspect ratios are chosen 
similarly. Similar behavior has 
earlier been seen in simulations in Cartesian domains with 
aspect ratios not equal to unity \citep{BDS02}.
What is particularly interesting in our case is that,
in both Cartesian and spherical coordinates systems,  
the observed cell structures are persistent,
with lifetimes larger than the duration of our longest 
simulations which, in turn, are
longer than the magnetic diffusion time based on   
$L_\phi$ (for spherical runs) or $L_z$ (for Cartesian runs).
For example, we have checked
that in the case of run {\tt S7} the cell structures
remain unchanged for at least as long as 400 dissipative times, $\teta$.

\section{Conclusion}
\label{conclusion}
We have made a detailed numerical study of the effects of
changes in the geometrical shape and size of 
the spherical wedge domains on the growth and 
saturation of large-scale magnetic fields.
We have used direct three-dimensional numerical simulations of 
helically forced MHD equations 
using random helical forcing. 

For the smallest domain with aspect ratio close to one
we find dynamo action resulting in magnetic fields on scales larger
than the characteristic scale of the forcing. The large-scale magnetic energy 
grows to exceed the kinetic energy over diffusive time scales, similar to 
that seen earlier in Cartesian simulations in periodic boxes.

In domains larger in the azimuthal direction the large-scale magnetic field 
organizes itself in cell-like structures in the azimuthal direction.
The aspect ratio of the individual cells is close to unity.  
This large-scale pattern in the azimuthal direction has $m=8$.
This is determined by the smallest (radial) scale in the simulations.
[As an aside, somewhat similar behavior was found by \cite{MTB90}
in a study of mean-field dynamos.]
To encompass such a structure within the computational domain
we have to have $L_{\phi}\ge(\pi/4)r_2$. Further increases in the size of the 
domain in the azimuthal direction just make the cells repeat themselves,
tiling the domain along the azimuthal direction, and thus resulting
in very small longitudinally averaged fields for larger domain sizes.
We note that this implies that extrapolation of results from Cartesian box
simulations with unit aspect ratio to spherical
shells can be misleading. We have also studied the 
effects of increasing the size of the domain in the meridional extent.  
The resulting large-scale magnetic fields show little qualitative change 
except for a marginal increase, provided $L_{\theta}\ge(\pi/5)r_2$;
see Fig.~\ref{phi-largescale}. Hence the smallest wedge shaped domain
in which we can expect to observe features of simulations in a full sphere
must have $L_{\phi}=(\pi/4)r_2$ and $L_{\theta}=(\pi/5)r_2$.

Furthermore the presence of the cellular structures along the 
azimuthal direction means that the usual employment of longitudinal 
averaging loses much information if used as a way to define 
large-scale magnetic fields.
Clearly a possible alternative is to define the large-scale magnetic field
via Fourier transform along the $\phi$ direction of our domain.
A large-scale magnetic energy defined in this fashion results
in strong fields of equipartition strength in all the domain sizes we have
used. 
We note here that the large-scale magnetic field defined using Fourier
filtering obeys some of the the Reynolds rules only approximately.
For example the average of the product of an average and a fluctuation
vanishes only for infinite scale separation.
This shortcoming may cause some discrepancies between theory and model,
which is however beyond the scope of the present paper.

In all our simulations we find the cellular structures to be long-lived
with lifetimes longer than the duration of our simulations. 
This therefore suggests that these structures are not transients.
In an attempt to compare with mean-field dynamo models one 
must note that the excitation conditions for modes with $m>2$ are normally 
much
higher than for $m$ below 2, although there is a clear trend for
this difference to diminish for thinner shells \citep{BTR89}.
On the other hand, anisotropies of the  $\alpha$ effect might
significantly change this.

It is important to clarify the similarities and differences
between our results and those of previous studies.
Forced turbulence simulations have been carried out by
\cite{MM06,MMT07} who also adopted a forcing function
in terms of Chandrasekhar--Kendall functions -- although not random -- 
and used perfectly conducting boundary conditions.
They included the effects of rotation and considered both
helical and non-helical forcings.
Their computational domain is a full sphere.
They considered laminar flow patterns and found large-scale
magnetic fields to be generated, 
but the energy contained in the large-scale component
is generally small compared with the kinetic energy.
Fully turbulent simulations in spherical shells have been
studied by \cite{BT02,BMT04,BMT06,BBBMNT08}.
These flows are subject to rotation and stratification which make
them helical.
However, the degree of helicity is weak compared to our fully helical
forcing functions and a broad range of wavenumbers is being driven,
so it is difficult to identify a well-defined energy-carrying scale.

Our simulations show the effects of magnetic helicity conservation
(see Fig.~\ref{bsat}),
but the magnetic Reynolds number is still
rather low, so it may be of interest to repeat such simulations at
larger magnetic Reynolds numbers.
However, it is important to run for sufficiently long times to
be able to obtain full saturation.
Obviously, such long saturation times are not astrophysically relevant,
and earlier work in Cartesian domains gives clear predictions that the
constraints from magnetic helicity are alleviated in the presence of
shear giving rise to small-scale magnetic helicity fluxes \citep{B05,KKB08}.
Allowing for latitudinal differential shear motions is therefore one of
our next objectives.

\acknowledgements
Dhrubaditya Mitra is supported by the Leverhulme Trust.
He and RT thank Nordita for hospitality during the
program `Turbulence and Dynamos'.
AB and David Moss thank the Astronomy Unit, 
Queen Mary University of London, for hospitality. 
Computational resources were granted by
UKMHD, QMUL HPC facilities purchased under the SRIF initiative,
and the National Supercomputer Centre in Link\"oping in Sweden.

\appendix
\section{Random helical forcing in spherical coordinates}
\label{helf}
In this Appendix we briefly describe the helical forcing used 
in our simulations in spherical wedge 
domains. We require the helicity of the forcing to be positive at every time-step
at every grid point. Furthermore in order to obtain a turbulent state we 
use random forcing which is white-in-time. In Cartesian coordinates this is 
achieved by using appropriately normalized Beltrami waves~\citep{B01};
 in the spherical case we need to use the Chandrasekhar-Kendall 
function~\citep{CK57}. 
Similar forcing functions, although not random, in spherical coordinate systems 
have also been discussed by \cite{LHT07}.

To guarantee positive helicity we demand, following \cite{CK57},
\begin{equation}
\curl\f = \alpha \f
\label{eigencurl}
\end{equation}
with a positive $\alpha$ at every point in our computational domain. This in turn 
implies that $\f$ should have the form 
\begin{equation}
\curl\curl\f = \alpha^2\f, 
\end{equation}
which, using $\dive\f=0$, becomes
\begin{equation}
\lap \f + \alpha^2\f = 0 .
\label{feq}
\end{equation}
Clearly all solutions of this equation are solutions of Eq.~(\ref{eigencurl})
but the converse is not true. To find solutions of
(\ref{feq}) consider a scalar function $\psi$ satisfying 
the Helmholtz equation, 
\begin{equation}
\lap\psi+\alpha^2\psi=0 ,
\label{helmholtz}
\end{equation}
whose solutions in spherical polar coordinates are obtained in
terms of spherical Bessel function and spherical harmonics, 
\begin{equation}
\psi=\sum_{l=0}^{\infty}\sum_{m=-l}^{l}
      z_l(\alpha r)Y^m_l(\theta,\phi)\exp(\i\xi_m),
\label{solhelmoholtz}
\end{equation}
where
\begin{equation}
z_l(\alpha r)=a_lj_l(\alpha r) + b_ln_l(\alpha r).
\end{equation}
Here $j_l$ and $n_l$ are spherical Bessel functions of the first and second kind
respectively and $a_l$ and $b_l$ are constants determined by the boundary conditions. 
A solution of Eq.~(\ref{eigencurl}) can then be constructed as the sum 
\begin{equation}
\f = {\bf T} + {\bf S},
\end{equation}
where
\begin{equation}
{\bf T} = \curl({\bf e}\psi), ~~~~\/ {\bf S} = \frac{1}{\alpha}\curl{\bf T}.
\label{helical_force}
\end{equation}
We wish to confine our forcing to certain bands of length scales and also to
randomize it. The characteristic scales of the forcing function in the radial,
meridional and azimuthal direction are given by $\alpha$, $l$ and $m$ respectively. 
As to the choice of boundary conditions, we demand that $\f$ is zero at the 
two radial boundaries $r=r_1$ and $r=r_2$. 
The constants $a_l$, $b_l$ and $\alpha$ are then related by 
\begin{equation}
a_lj_l(\alpha r_1) + b_ln_l(\alpha r_1) = a_lj_l(\alpha r_2) + b_ln_l(\alpha r_2) = 0.
\label{alphaeqn}
\end{equation}
For a particular choice of $l$ this transcendental equation has an infinite number of 
solutions for $\alpha$ and the ratio $a_l/b_l$. A higher value of $\alpha$ implies more 
zeros of the function $z_l(\alpha r)$ lies within $r_1$ and $r_2$, which in turn
implies that the characteristic radial scale of $z_l(\alpha r)$ becomes smaller.
Note that we have periodic boundary conditions along the azimuthal direction,
hence the non-zero values of $m$ which are possible in our domain depends on the
extent of the domain in the azimuthal direction, i.e., 
$m_{\rm min}= 2\pi/L_{\phi}$, e.g., $m_{\rm min}=20$ for the run {\tt S1}. In order to 
mimic turbulence, we force at the intermediate length scales which allows 
kinetic
energy to cascade to smaller scales. Furthermore, we want the forcing to 
go to zero at the equator. This implies $l$ must be odd. 
The values of $\alpha$, $l$ and $m$ that we use for the run {\tt S1} 
are given in Table~\ref{table_alpha}. We used the GNU scientific library 
\footnote{\url{http://www.gnu.org/software/gsl/}.}
to compute the Bessel functions and spherical harmonics in our code.  
\begin{deluxetable}{ccccc}
\tablecaption{Values of $\alpha$ that satisfy Eq.~(\ref{alphaeqn}) used in 
the run {\tt S1}. 
\label{table_alpha}}
\tablehead{
\colhead{$m$} & \colhead{$l$} & \colhead{$\alpha_1$} & 
             \colhead{$\alpha_2$} & \colhead{$\alpha_3$}}
\startdata
20 & 81 & 129.011139 & 135.938721 & 143.325378 \\ 
20 & 83 & 130.880829 & 137.703308 & 144.992371 \\ 
20 & 85 & 132.771484 & 139.489746 & 146.681885 \\ 
20 & 87 & 134.682465 & 141.297455 & 148.393219 \\ 
20 & 89 & 136.613068 & 143.125763 & 150.125793 \\ 
40 & 81 & 129.011139 & 135.938721 & 143.325378 \\ 
40 & 83 & 130.880829 & 137.703308 & 144.992371 \\ 
40 & 85 & 132.771484 & 139.489746 & 146.681885 \\ 
40 & 87 & 134.682465 & 141.297455 & 148.393219 \\ 
40 & 89 & 136.613068 & 143.125763 & 150.125793 \\ 
60 & 91 & 138.562683 & 144.974060 & 151.879028 \\ 
60 & 93 & 140.530670 & 146.841827 & 153.652344 \\ 
60 & 95 & 142.516434 & 148.728455 & 155.445251 \\ 
60 & 97 & 144.519348 & 150.633453 & 157.257141 \\ 
60 & 99 & 146.538788 & 152.556305 & 159.087616 \\ 
80 & 121 & 169.644516 & 174.748535 & 180.321686 \\ 
80 & 123 & 171.805023 & 176.847809 & 182.341858 \\ 
80 & 125 & 173.971619 & 178.958252 & 184.375427 \\ 
80 & 127 & 176.143417 & 181.079208 & 186.421967 \\ 
80 & 129 & 178.319519 & 183.209961 & 188.481140 \\ 
100 & 121 & 169.644516 & 174.748535 & 180.321686 \\ 
100 & 123 & 171.805023 & 176.847809 & 182.341858 \\ 
100 & 125 & 173.971619 & 178.958252 & 184.375427 \\ 
100 & 127 & 176.143417 & 181.079208 & 186.421967 \\ 
100 & 129 & 178.319519 & 183.209961 & 188.481140 \\ 
\enddata
\end{deluxetable}

\section{The Pencil Code in spherical polar coordinates}
\label{noncart}
The \textsc{Pencil Code} was originally written in 
Cartesian coordinates. To use it for our simulations
of the compressible MHD equations in spherical polar coordinates,
it needs to be changed accordingly. In fact, given its
modularity, the \textsc{Pencil Code} is well suited to be
generalized to any curvilinear coordinate system. 
We do this by writing the MHD equations in a covariant
form by replacing partial derivatives by covariant derivatives.
We shall illustrate this method by considering the particular
case of spherical coordinates, which is the one
relevant to our simulations here.
Let us first consider the divergence of a vector field $\AAA$.
In Cartesian coordinates using index notation
\begin{equation}
\dive\AAA = A_{\alpha,\alpha},
\label{divA}
\end{equation}
where a comma denotes partial differentiation.
The same operator can be written in any non-Cartesian coordinate 
system by replacing the partial derivative by the covariant derivative
denoted by a semicolon thus:
\begin{equation}
A^{\alpha}_{;\beta} \equiv A^{\alpha}{,\beta} - \Gamma^{\alpha}_{\sigma \beta}A_{\sigma} 
\label{covariant}
\end{equation}
where $\Gamma^{\alpha}_{\sigma\beta}$ are the connection coefficients 
obtained from the metric corresponding to the coordinates chosen. In the case of 
spherical coordinates the metric takes the form
\begin{equation}
g_{\alpha \beta} = \begin{pmatrix}
1 & 0 & 0 \cr
0 & r^{-1} & 0 \cr
0 & 0 & (r\sin\theta)^{-1}
\end{pmatrix} .
\end{equation}
We shall write the covariant derivatives in the non-coordinate bases,
by defining a new triplet of coordinate differentials
\begin{equation}
\dd{\hat r} = \dd r,\quad
\dd{\hat \theta}=r\,\dd{\theta},\quad\mbox{and}\quad
\dd{\hat \phi}=r\sin\theta\,\dd{\phi} .
\end{equation}
In these bases the connection coefficients take a particularly simple form,
\begin{equation}
{\Gamma^{\hat\theta}}_{{\hat r}{\hat\theta}}
={\Gamma^{\hat\phi}}_{{\hat r}{\hat\phi}}
=-{\Gamma^{\hat r}}_{{\hat\theta}{\hat\theta}}
=-{\Gamma^{\hat r}}_{{\hat\phi}{\hat\phi}}
=1/r,
\end{equation}
\begin{equation}
{\Gamma^{\hat\phi}}_{{\hat\theta}{\hat\phi}}
=-{\Gamma^{\hat\theta}}_{{\hat\phi}{\hat\phi}}
=\cot\theta/r,
\end{equation}
with all other connection coefficients being zero. 
This simplification makes this non-coordinate
basis particularly appealing for numerical simulations. 
For example, for the divergence of a
vector $\AAA$ we obtain
\begin{equation}
\dive\AAA={A_{\hat\alpha; \hat\alpha}}
         = A_{\hat\alpha,\hat\alpha}
+2r^{-1}A_{\hat r}+r^{-1}\!\cot\!\theta A_{\hat\theta},
\label{divAA}
\end{equation}
where
\begin{equation}
A_{\hat\alpha,\hat\alpha} = \partial_r A_{\hat r} + \frac{1}{r}\partial_{\theta} A_{\hat \theta}
                            +  \frac{1}{r\sin\theta}\partial{\phi}A_{\hat \phi} .
\end{equation}
Note that in the non-coordinate basis 
the metric tensor is the Kronecker delta and so
the covariant and contravariant components
of a tensor are one and the same, hence in the above expression we have not 
distinguished between them. 
As in Eq.~(\ref{divAA}) any vector differential 
operator in curvilinear coordinate system can be written 
as the sum of two parts: the first involving the vector operator in the 
Cartesian
form with added scaling factors $r^{-1}$ and $(r\sin\theta)^{-1}$,
and the other part involving the connection coefficients.
The modular feature of the \textsc{Pencil Code} then plays an
important role since the derivatives in \textsc{Pencil Code} are computed in a separate module,
and all we need to do to adapt the \textsc{Pencil Code} to any non-Cartesian coordinate system
is to change this derivative module by adding the scaling factors corresponding
to the coordinate system chosen. The vector operators, e.g., divergence, curl, Laplacian etc,
are then calculated in a different module which uses the derivative module. The parts
which depend on the connection coefficients are added to this module.
The other minor changes to the code involves coding new boundary conditions
and new modules to calculate volume averages.
All these changes are now part of the public release  of the code.

For completeness, we list here the expressions for the most commonly used
vector differential operators in our code.
For the curl of a vector field $\AAA$ we have
\begin{equation}
\curl\AAA=
\begin{pmatrix}
A_{\hat\phi;\hat\theta}-A_{\hat\theta;\hat\phi}\cr
A_{\hat r;\hat\phi}-A_{\hat\phi;\hat r}\cr
A_{\hat\theta;\hat r}-A_{\hat r;\hat\theta}
\end{pmatrix}
=
\begin{pmatrix}
A_{\hat\phi,\hat\theta}-A_{\hat\theta,\hat\phi}\cr
A_{\hat r,\hat\phi}-A_{\hat\phi,\hat r}\cr
A_{\hat\theta,\hat r}-A_{\hat r,\hat\theta}
\end{pmatrix}+
\begin{pmatrix}
r^{-1}\!\cot\!\theta A_{\hat \phi}\cr -r^{-1}A_{\hat \phi}\cr r^{-1}A_{\hat \theta} \cr
\end{pmatrix}.
\end{equation}
For the advective operator we obtain
\begin{eqnarray}
({\bf u}\cdot{\bf \nabla}\AAA)_{\hat r}
&=&u_{\hat r}A_{\hat r,\hat r}+u_{\hat\theta}A_{\hat r,\hat\theta}+u_{\hat\phi}A_{\hat r,\hat\phi}
-r^{-1}\,u_{\hat\theta}A_{\hat\theta}-r^{-1}\,u_{\hat\phi}A_{\hat\phi}
\cr
({\bf u}\cdot{\bf \nabla}\AAA)_{\hat\theta}
&=&u_{\hat r}A_{\hat\theta,\hat r}+u_{\hat\theta}A_{\hat\theta,\hat\theta}+u_{\hat\phi}A_{\hat\theta,\hat\phi}
+r^{-1}\,u_{\hat\theta}A_{\hat r}-r^{-1}\!\cot\!\theta\,u_{\hat\phi}A_{\hat\phi}
\cr
({\bf u}\cdot{\bf \nabla}\AAA)_{\hat\phi}
&=&u_{\hat r}A_{\hat\phi,\hat r}+u_{\hat\theta}A_{\hat\phi,\hat\theta}+u_{\hat\phi}A_{\hat\phi,\hat\phi}
+r^{-1}\,u_{\hat\phi}A_{\hat r}+r^{-1}\!\cot\!\theta\,u_{\hat\phi}A_{\hat\theta}.
\end{eqnarray}
To calculate the second order differential operators we need the expression for second
order covariant derivative given by 
\begin{eqnarray}
A_{\hat\alpha;\hat\beta\hat\gamma}
&=&A_{\hat\alpha;\hat\beta,\hat\gamma}
-{\Gamma^{\hat\sigma}}_{\hat\alpha\hat\gamma}\,A_{\hat\sigma;\hat\beta}
-{\Gamma^{\hat\sigma}}_{\hat\beta\hat\gamma}\,A_{\hat\alpha;\hat\sigma}
\cr
&=&A_{\hat\alpha,\hat\beta\hat\gamma}
-{\Gamma^{\hat\sigma}}_{\hat\alpha\hat\beta}\,A_{\hat\sigma,\hat\gamma}
-{\Gamma^{\hat\sigma}}_{\hat\alpha\hat\beta,\hat\gamma}\,A_{\hat\sigma}
-{\Gamma^{\hat\sigma}}_{\hat\alpha\hat\gamma}\,A_{\hat\sigma,\hat\beta}
+{\Gamma^{\hat\sigma}}_{\hat\alpha\hat\gamma}{\Gamma^{\hat\nu}}_{\hat\sigma\hat\beta}\,A_{\hat\nu}
-{\Gamma^{\hat\sigma}}_{\hat\beta\hat\gamma}\,A_{\hat\alpha,\hat\sigma}
+{\Gamma^{\hat\sigma}}_{\hat\beta\hat\gamma}{\Gamma^{\hat\nu}}_{\hat\alpha\hat\sigma}\,A_{\hat\nu}.
\end{eqnarray}
For example the Laplacian of a scalar field $\Psi$ is given by
\begin{equation}
\Delta\Psi = E_{\hat\beta;\hat\beta}
= (\partial_{\hat\beta}\Psi)_{,\hat\beta} + 
  \frac{2}{r}\Psi_{,\hat r} + \frac{\cot\!\theta}{r}\Psi_{,\hat\theta} ,
\end{equation}
and the $\mbox{grad}\,\mbox{div}$ operator takes the form
\begin{eqnarray}
\grad\dive\AAA =
\begin{pmatrix}
                A_{\hat\alpha,\hat\alpha\hat r}+2r^{-1}\,A_{\hat r,\hat r}
             +r^{-1}\cot\!\theta A_{\hat\theta,\hat r}-2r^{-2}\,A_{\hat r}
             -r^{-2}\cot\!\theta A_{\hat\theta}\cr
            A_{\hat\alpha,\hat\alpha\hat\theta}+2r^{-1}\,A_{\hat r,\hat\theta}
             +r^{-1}\cot\!\theta A_{\hat\theta,\hat\theta}
             -r^{-2}\sin\!^{-2}\theta A_{\hat\theta}\cr
           A_{\hat\alpha,\hat\alpha\hat\phi}+2r^{-1}A_{\hat r,\hat\phi}
             +r^{-1}\cot\!\theta A_{\hat\theta,\hat\phi}
\end{pmatrix}.
\end{eqnarray}

\bibliographystyle{aa}
\bibliography{sunref}

\begin{thebibliography}{100}
\expandafter\ifx\csname natexlab\endcsname\relax\def\natexlab#1{#1}\fi

\bibitem[{Brandenburg(2001)}]{B01}
Brandenburg, A. 2001, ApJ, 550, 824

\bibitem[{Brandenburg(2005)}]{B05}
Brandenburg, A. 2005, ApJ, 625, 539

\bibitem[{Brandenburg \& Dobler (2001)}]{BD01}
Brandenburg, A. \& Dobler, W. 2001, A\&A, 369, 329


\bibitem[{Brandenburg \& Dobler (2002)}]{BD02b}
Brandenburg, A. \& Dobler, W. 2002, Comp. Phys. Comm. 147, 471


\bibitem[{Brandenburg, Dobler, \& Subramanian (2002)}]{BDS02}
Brandenburg, A., Dobler, W., \& Subramanian, K., 2002, AN, 323, 99

\bibitem[{Brandenburg, Tuominen, \& R\"adler (1989)}]{BTR89}
Brandenburg, A., Tuominen, T., \& R\"adler, K. H., 1989, GAFD, 49, 45

\bibitem[{Brown et al.\ (2007)}]{BBBMNT08}
Brown, B. P., Browning, M. K., Brun, A. S., Miesch, M. S., Nelson, N. J., \&
Toomre, J., 2007, AIPC, 948, 271

\bibitem[{Browning (2008)}]{BR08} Browning, M.~K., 2008, ApJ, 676, 1262


\bibitem[{Brun et al.\ (2002)}] {BT02}
Brun, A. S., Toomre, J. 2002, ApJ, 570, 865

\bibitem[{Brun et al.\ (2004)}] {BMT04}
Brun, A. S., Miesch, M. S., Toomre, J. 2004, ApJ 614, 1073

\bibitem[{Brun et al.\ (2006)Brun, Miesch, \& Toomre}]{BMT06}
Brun, A.S., Miesch, M., \& Toomre, J. 2006, ApJ, 614, 1073

\bibitem[{Chandrasekhar \& Kendall(1957)}]{CK57}
Chandrasekhar, S. \& Kendall, P. 1957, Astrophys.~J., 126, 457

\bibitem[{Gilman (1983)}]{G83}
Gilman, P.~A., 1983, ApJS, 53, 243

\bibitem[{Gilman \& Glatzmaier(1981)}]{GG81}
Gilman, P.~A. \& Glatzmaier, G.~A. 1981, ApJS, 45, 335

\bibitem[{Gilman \& Miller (1981)}]{GM81}
Gilman, P.~A. \& Miller, J., 1981, ApJS, 46, 211

\bibitem[{Glatzmaier (1984)}] {G84}
Glatzmaier, G.A., 1984, J. Comp. Phys., 55, 461 

\bibitem[{Glatzmaier (1985)}] {G85} Glatzmaier, G.A., 1985, ApJ, 291, 300 

\bibitem[{Glatzmaier \& Gilman (1982)}]{GG82}
Glatzmaier, G.~A. \& Gilman, P.~A, 1982, GAFD, 31, 137

\bibitem[{Hughes \& Proctor (2008)}] {HP08}
Hughes, D. W., \& Proctor, M. R. E. 2008, Phys. Rev. Lett., 102, 044501

\bibitem[{K\"apyl\"a et al.\ (2008)}] {KKB08}
K\"apyl\"a, P. J., Korpi, M. J., \& Brandenburg, A., 2008, A\&A, 491, 353

\bibitem[{Krause \& Steenbeck (1967)}] {KS67}
Krause, F., \& Steenbeck, M. (1967), Z.\ Naturforsch., 22a, 671

\bibitem[{Livermore, Hughes \& Tobias (2007)}] {LHT07}
Livermore, P.W., Hughes, D.W. \& Tobias, S.M. 2007, Phys. Fluids 19, 057101 

\bibitem[{Miesch et al.\ (2000)}] {MET00}
Miesch, M. S., Elliott, J. R., Toomre, J., et al.\ 2000, ApJ, 532, 593

\bibitem[{Mininni \& Montgomery (2006)}] {MM06}
Mininni, P. D., \& Montgomery, D. C. 2006, Phys. Fl. 18, 116602

\bibitem[{Mininni, Montgomery, \& Turner (2007)}] {MMT07}
Mininni, P. D., Montgomery, D. C., \& Turner, L. 2007, New J. Phys. 9, 303

\bibitem[{Moss, Tuominen \& Brandenburg (1990)}] {MTB90}
Moss, D., Tuominen, I., \& Brandenburg, A., 1990, A\&A 240, 142

\end{thebibliography}
\end{document}